\documentclass{ws-p10x7}

\newcommand{\DO}{D\raise1pt\hbox{$\not$}O}

\begin{document}

\title{Inclusive Direct-photon and $\pi^0$ production in proton--nucleon 
collisions}

\author{T.~Ferbel}

\address{University of Rochester, Rochester, New York 14627}         

\twocolumn[\maketitle\abstract{
We present a study of inclusive direct-$\gamma$ and $\pi^0$ production in
hadronic interactions that focuses on a comparison of the ratio of
$\gamma$ and $\pi^0$ yields with expectations from next-to-leading order
perturbative QCD (NLO pQCD).}] 

\section{Introduction and the $\gamma / \pi^0$ Ratio}

Direct-photon production in hadronic collisions at high transverse
momenta ($p_T$) has long been viewed as an ideal vehicle for extracting 
the gluon content, $G(x)$, of hadrons. The sensitivity to $G(x)$ arises 
from the contribution of the Compton subprocess $gq\rightarrow\gamma q$
to direct-$\gamma$ production. $G(x)$ is well constrained by other data
for $x<0.25$, but not at larger~$x$. In principle, a precise
measurement of direct-photon
production at fixed-target energies can constrain $G(x)$ at large~$x$, 
and such data have been used in fits to global parton distribution
functions (PDF).

\begin{figure}
\epsfxsize=2.5truein
\vspace{-0.5in}
\centerline{\epsffile{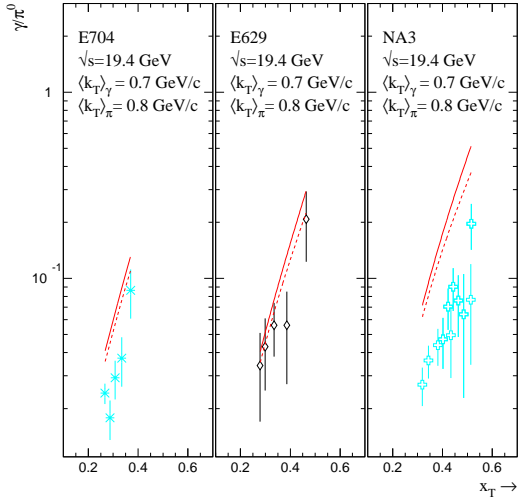}}
\vspace{-0.5in}
\caption{Comparison of $\gamma/\pi^0$ rates 
as a function of $x_T$ for NA3, E704, and E629 at $\sqrt{s}=19.4$~GeV.
Overlayed are the results from NLO pQCD (solid) and $k_T$-enhanced
calculations (dashed).  Values of $\langle k_T\rangle$ used for the
$k_T$-enhanced calculations are given in the legend.
\label{fig:photonTOpi0_1}}
\end{figure}

Deviations have been noted between measured 
inclusive direct-photon cross sections and NLO pQCD.    
Ratios of data to theory for both $\gamma$ 
and pion production  indicate substantial disagreement between 
data and pQCD, as well as between different experiments.\cite{phenom} 
The latter is not completely
surprising, because, especially for direct-photon production, signals 
are often small and backgrounds large, especially at lower energies. 
Several experiments show
better agreement with NLO pQCD than others, but the results do not
provide confidence in the theory nor in the quality of all 
data.  Although it has been suggested that deviations from theory for
both $\gamma$ and pion production can be ascribed to higher-order effects of
initial-state soft-gluon radiation, it seems unlikely that theoretical 
developments alone will accommodate the observed level of scatter
in data/theory. These discrepancies motivated us to consider 
measurements of the $\gamma/\pi^0$ ratio over the available range of 
center-of-mass energies ($\sqrt{s}$).  
Both experimental and theoretical 
uncertainties tend to cancel in such a ratio, and it is 
also less sensitive to the treatment of gluon radiation. 
A sample of the ratio of direct-$\gamma$ to $\pi^0$ cross sections for 
both data and NLO pQCD is given for incident protons, as a function 
of $x_T=2p_T/\sqrt{s}$, in 
Figs.~\ref{fig:photonTOpi0_1}--\ref{fig:photonTOpi0_3}.  
The results at $\sqrt{s}=19.4$~GeV 
are displayed in Fig.~\ref{fig:photonTOpi0_1}. For all measurements, 
theory is high compared to data.~(The NLO calculations 
use a single scale of $\mu=p_T/2$, 
CTEQ4M PDFs,\cite{cteq4} and BKK fragmentation functions\cite{BKK} for 
pions.) Figure~\ref{fig:photonTOpi0_2} shows 
the~$\gamma$ to~$\pi^0$ ratio at $\sqrt{s}\approx23-24$~GeV.  Just 
as in Fig.~\ref{fig:photonTOpi0_1}, theory is high relative to data.
At larger $\sqrt{s}$, the NLO value for the ratio agrees better with 
experiment, as seen in Fig.~\ref{fig:photonTOpi0_3} for $\sqrt{s}=31-39$~GeV.
At even higher $\sqrt{s}$, theory lies slightly below the data.\cite{phenom} 
A compilation of these results, displayed for 
simplicity without their uncertainties, is presented in 
Fig.~\ref{fig:photonTOpi0_all}.  
Here, the ratio of data to theory was fitted to a constant value at 
high-$p_T$, and the results plotted as a function of $\sqrt{s}$.  The 
results suggest an energy dependence in the ratio of data to theory for 
$\gamma/\pi^0$ production, already noted in 
Figs.~\ref{fig:photonTOpi0_1}--\ref{fig:photonTOpi0_3}.  There are also  
substantial differences between experiments at low $\sqrt{s}$, where the
observed $\gamma/\pi^0$ is smallest, which makes it difficult to
quantify this trend.  Recognizing the presence of these differences is
especially important because only the direct-photon experiments at low 
energy have been used in PDF fits to $G(x)$.

\section{Corrections for Soft Gluon Emission}

\begin{figure}
\epsfxsize=2.5truein
\vspace{-0.5in}
\centerline{\epsffile{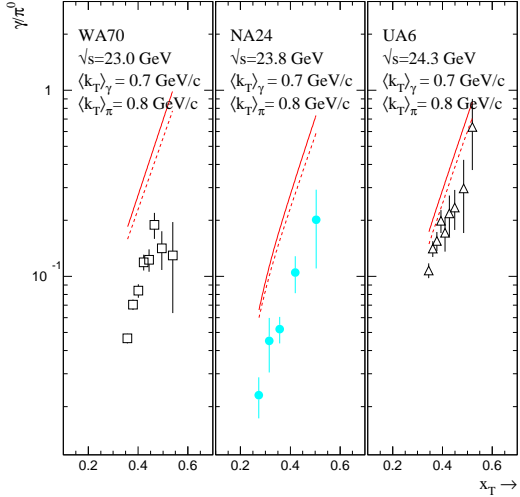}}
\vspace{-0.5in}
\caption{Comparison of $\gamma/\pi^0$ rates 
as a function of $x_T$ for WA70, NA24, and UA6 at
$\sqrt{s}\approx23-24$~GeV. (See text and Fig. 1 for additional explanation.)
\label{fig:photonTOpi0_2}}
\end{figure}

\begin{figure}
\epsfxsize=2.5truein
\vspace{-0.5in}
\centerline{\epsffile{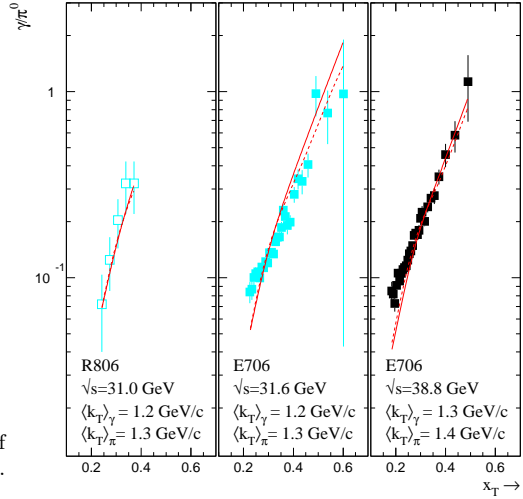}}
\vspace{-0.5in}
\caption{Comparison of $\gamma/\pi^0$ rates as a function of $x_T$ for 
R806 at $\sqrt{s}=31$~GeV and E706 at $\sqrt{s}=31.6$ and $38.8$~GeV.
(See text and Fig. 1 for additional explanation.)
\label{fig:photonTOpi0_3}}
\end{figure}

\begin{figure}
\epsfxsize=2.5truein
\vspace{-0.5in}
\centerline{\epsffile{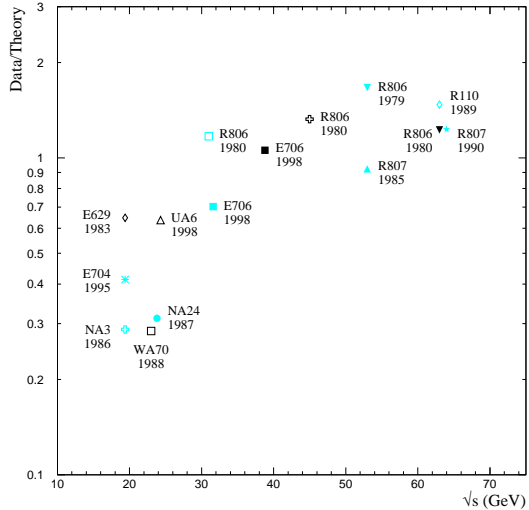}}
\vspace{-0.5in}
\caption{Ratio of data to NLO for the $\gamma/\pi^0$ production 
as a function of $\sqrt{s}$.\protect \cite{phenom} The values represent fits to
the ratio of data to NLO pQCD theory, without $k_T$-enhancement (see
text).
\label{fig:photonTOpi0_all}}
\end{figure}

In the absence of a rigorous theoretical treatment, 
a more intuitive, but often successful, phenomenological approach 
has been used to describe soft gluon radiation in high-$p_T$ inclusive 
production,\cite{ktprd} and parametrized in 
terms of an effective $k_T$ that provided  additional 
transverse impulse to the outgoing partons. This provided
$p_T$-dependent corrections to the NLO pQCD calculations.  
The corrections for direct-$\gamma$ and $\pi^0$ production in Fermilab 
experiment E706 are large (and comparable) over the full range of 
$p_T$.\cite{phenom} The corrections depend on the
values used for $\langle k_T\rangle$, with changes of 200 MeV/$c$ making
substantial difference, and therefore making it difficult to obtain the 
precision needed for extracting global parton distributions.  
In addition, there are different ways to implement such 
models,\cite{phenom,mrst} 
which can produce quantitative differences in the $k_T$-correction factors. 
However, it is expected that, 
in the ratio of $\gamma$ and $\pi^0$ cross sections, the impact of $k_T$ 
corrections should be minimal, and this is observed in
Figs.~\ref{fig:photonTOpi0_1}--\ref{fig:photonTOpi0_3}, where the
dashed curves indicate the predicted ratios using previous $k_T$
corrections.\cite{ktprd} ~Thus, it seems that the trend  in
Fig.~\ref{fig:photonTOpi0_all} cannot be understood purely on the
basis of corrections for $k_T$.

Resummed pQCD calculations for single direct-photon production are 
currently under 
development.\cite{nason,kidonakisowens,laenen,lilai,li,sterman} 
Two recent threshold-resummed pQCD calculations 
for direct photons\cite{nason,kidonakisowens} exhibit far less dependence 
on QCD scales than NLO theory, and provide an
enhancement at high~$p_T$.
A method for simultaneous treatment of recoil and threshold
corrections in inclusive single-$\gamma$ cross sections is also being
developed.\cite{sterman} This approach accounts explicitly for the 
recoil from soft radiation in the hard-scattering, and 
conserves both energy and transverse momentum for the resummed radiation.  
The possibility of substantial enhancements from higher-order perturbative
and power-law nonperturbative corrections relative to NLO are
indicated at both moderate and high $p_T$ for fixed-target energies,
similar to the enhancements obtained with simple 
$k_T$-smearing.\cite{ktprd} 

Although there are discrepancies between experiments, especially significant 
at low $\sqrt{s}$, there appears to be an unexplained systematic 
trend with energy in Fig.~\ref{fig:photonTOpi0_all}. Hopefully, this can
be clarified once resummation calculations for inclusive pion production 
become available. Nevertheless, the recent developments in theory of 
direct-$\gamma$ processes 
provide cause for optimism that the 
long-standing difficulties in developing an adequate 
description of direct-$\gamma$ production can eventually be resolved, 
making possible a reliable extraction of $G(x)$ from such data.

\medskip
This report is based on work performed with L. Apanasevich, 
M.~Begel, C. Bromberg, G. Ginther, J. Huston, S. Kuhlmann, P. Slattery, M.
Zielinski, and V. Zutshi. I also wish to thank P.~Aurenche, D. Soper, and
G.~Sterman for helpful discussions.

% FFFFFFFFFFFFFFFFFFFFFFFFFFFFFFFFFFFFFFFFFF

%\bibliography{paper}  
%\bibliographystyle{prsty}
%
% Figures
%
%

%\newpage

%\newpage

%\newpage
\end{document}